\documentclass[twocolumn]{whitepaper}

\title{New Observations Needed to Advance Our Understanding of Coronal Mass Ejections}

\author[1]{Erika~Palmerio~\orcidlink{0000-0001-6590-3479}}
\author[2]{Benjamin~J.~Lynch~\orcidlink{0000-0001-6886-855X}}
\author[2]{Christina~O.~Lee~\orcidlink{0000-0002-1604-3326}}
\author[3]{Lan~K.~Jian~\orcidlink{0000-0002-6849-5527}}
\author[3]{Teresa~Nieves-Chinchilla~\orcidlink{0000-0003-0565-4890}}
\author[4]{Emma~E.~Davies~\orcidlink{0000-0001-9992-8471}}
\author[5]{Brian~E.~Wood~\orcidlink{0000-0002-4998-0893}}
\author[4]{No{\'e}~Lugaz~\orcidlink{0000-0002-1890-6156}}
\author[4]{R{\'e}ka~M.~Winslow~\orcidlink{0000-0002-9276-9487}}
\author[1]{Tibor~T{\"o}r{\"o}k~\orcidlink{0000-0003-3843-3242}}
\author[4]{Nada~Al-Haddad~\orcidlink{0000-0002-0973-2027}}
\author[4]{Florian~Regnault~\orcidlink{0000-0002-4017-8415}}
\author[6]{Meng~Jin~\orcidlink{0000-0002-9672-3873}}
\author[4]{Camilla~Scolini~\orcidlink{0000-0002-5681-0526}}
\author[3,7]{Fernando~Carcaboso~\orcidlink{0000-0003-1758-6194}}
\author[4]{Charles~J.~Farrugia~\orcidlink{0000-0001-8780-0673}}
\author[1]{Vincent~E.~Ledvina~\orcidlink{0000-0003-0127-5105}}
\author[1]{Cooper~Downs~\orcidlink{0000-0003-1759-4354}}
\author[3,7]{Christina~Kay~\orcidlink{0000-0002-2827-6012}}
\author[3,8]{Sanchita~Pal~\orcidlink{0000-0002-6302-438X}}
\author[3,8]{Tarik~M.~Salman~\orcidlink{0000-0001-6813-5671}}
\author[9]{Robert~C.~Allen~\orcidlink{0000-0003-2079-5683}}

\affil[1]{Predictive Science Inc., San Diego, CA 92121, USA}
\affil[2]{Space Sciences Laboratory, University of California--Berkeley, Berkeley, CA 94720, USA}
\affil[3]{Heliophysics Science Division, NASA Goddard Space Flight Center, MD 20771, USA}
\affil[4]{Space Science Center, University of New Hampshire, Durham, NH 03824, USA}
\affil[5]{Naval Research Laboratory, Space Science Division, Washington, DC 20375, USA}
\affil[6]{Lockheed Martin Solar and Astrophysics Laboratory, Palo Alto, CA 94304, USA}
\affil[7]{The Catholic University of America, Washington, DC 20064, USA}
\affil[8]{Department of Physics and Astronomy, George Mason University, Fairfax, VA 22030, USA}
\affil[9]{Johns Hopkins University Applied Physics Laboratory, Laurel, MD 20723, USA}

\runningtitle{Observations to Advance Our Understanding of CMEs}
\shortauthors{Palmerio et al.}
\footer{White Paper submitted to the Heliophysics 2024--2033 Decadal Survey}

\begin{document}

\maketitle

\thispagestyle{firststyle}


\begin{abstract}

Coronal mass ejections (CMEs) are large eruptions from the Sun that propagate through the heliosphere after launch. Observational studies of these transient phenomena are usually based on 2D images of the Sun, corona, and heliosphere (remote-sensing data), as well as magnetic field, plasma, and particle samples along a 1D spacecraft trajectory (in-situ data). Given the large scales involved and the 3D nature of CMEs, such measurements are generally insufficient to build a comprehensive picture, especially in terms of local variations and overall geometry of the whole structure. This White Paper aims to address this issue by identifying the data sets and observational priorities that are needed to effectively advance our current understanding of the structure and evolution of CMEs, in both the remote-sensing and in-situ regimes. It also provides an outlook of possible missions and instruments that may yield significant improvements into the subject.

\end{abstract}


\section{Introduction}

Coronal mass ejections (CMEs) are among the most spectacular eruptions in the solar system, consisting of copious amounts of plasma and magnetic field that are regularly expelled from the Sun. After erupting and as they travel through interplanetary space, CMEs tend to expand and interact with the ambient solar wind, resulting in large structures \citep[measuring e.g.\ ${\sim}0.3$~AU in radial size by the time they reach 1~AU;][]{jian2018} that may have lost their twisted outer layers \citep{pal2021} and/or coherence \citep{owens2017} and are thus prone to deformations. The heliospheric evolution of CMEs may result in rotations, deflections, deformations, erosion, and re-configurations due to complex interactions of the ejected plasma with its surroundings \citep[e.g.,][]{manchester2017}, including with other CMEs \citep[e.g.,][]{lugaz2017}. These aspects make CMEs extremely complex to fully characterize in 3D, in terms of both their morphology and magnetic configuration, especially in light of the limited observations that have been historically available.

For example, CMEs were first observed through remote-sensing data from a single viewpoint, i.e.\ Earth, with early observations of the Sun and its corona---including transient phenomena such as CMEs---made in the early 70s with e.g.\ OSO-7, Skylab, and ground-based observatories. The following Solwind and SolarMax (in the late 70s--early 80s) as well as SOHO (in the 90s) satellites brought significant improvements in the temporal and spatial resolution of solar data, but it was not until the launch of the STEREO mission in the 2000s that the first remote-sensing images away from the Sun--Earth line would be taken. STEREO consisted of twin spacecraft, STEREO-A and STEREO-B (leading and trailing Earth in its orbit, respectively), equipped with solar disk, corona, and heliospheric imagers in their remote-sensing suite. The availability of multi-point observations of the Sun and its environment has led to major advances in CME research, including improving our understanding of CME morphology \citep[e.g.,][]{thernisien2009, wood2009}, better constraining CME propagation direction and speed through the solar corona as well as interplanetary space \citep[e.g.,][]{colaninno2013, mostl2015}, and enabling observations of CMEs that are ``stealthy'' from one viewpoint but evident from another \citep[e.g.,][]{nitta2021, palmerio2021b}. Currently, remote-sensing imagers away from the Sun--Earth line can be found onboard STEREO-A (STEREO-B was lost in 2014), as well as the more recently launched Parker Solar Probe (heliospheric imagers only) and Solar Orbiter (disk, corona, and heliospheric imagers). The latter two, however, do not have their remote-sensing instruments operational at all times, leaving STEREO-A a unique and persistent viewpoint to support solar imagery from near Earth.

On the other hand, in-situ measurements of the solar wind and interplanetary magnetic field---including transient phenomena such as the interplanetary counterparts of CMEs, also known as ICMEs---have been performed away from the Sun--Earth line already since the 60s and 70s, via mission programs such as Pioneer, Helios, and Voyager. However, given the large spatial scales involved, the likelihood of obtaining multi-point measurements of the same ICME is drastically reduced, so much so that the first analysis of the internal magnetic structure of CMEs was published only in the early 80s, using data from five different spacecraft between 1 and 2~AU \citep{burlaga1981}. Since then, multi-spacecraft studies of ICMEs have taken advantage of various heliophysics missions as well as planetary ones. These have enabled studies, among other topics, of the longitudinal variation \citep[e.g.,][]{farrugia2011, kilpua2011} and radial evolution \citep[e.g.,][]{good2019, salman2020} for selected ICME events. Nevertheless, such fortuitous spacecraft configurations are rather rare, and most analyses showcasing multi-point ICME measurements are characterized by arbitrary relative spacecraft geometries \citep[e.g.,][]{witasse2017, palmerio2021a}, which tend to complicate interpretation of an event, e.g.\ in discerning whether certain features are due to CME evolution in interplanetary space or to local distortions along the whole structure. Currently, in-situ measurements of the inner heliospheric environment are available from Earth, partially from Mars, as well as the STEREO-A, Parker Solar Probe, BepiColombo, and Solar Orbiter spacecraft. Given that each of these observers follows its own orbit around the Sun, studies that can address the internal structure and evolution of CMEs in interplanetary space via multi-point measurements are limited to those periods that are characterized by a propitious spacecraft configuration.

After over five decades of CME research (spanning over five solar cycles), we have obtained significant statistics on the characteristics and properties of CMEs from single-point (both remote-sensing and in-situ) observations \citep[see, e.g., the CME and ICME catalogs of][]{gopalswamy2009a, richardson2010, harrison2018, jian2018, nieveschinchilla2018}. Hence, it is clear that it is extremely challenging to reach a deeper insight on the intrinsic structure and evolution of CMEs based on a single viewpoint. To improve and expand our understanding in fundamental CME research, we need a set of dedicated observations that are aimed at treating CMEs as 3D structures that are in constant evolution and interaction with the ambient solar wind. In this White Paper, our goal is to identify such specific observations, in both the remote-sensing and in-situ regimes, and address their benefits for the studies of CMEs and ICMEs in a holistic way. We will also elaborate on possible missions that would be able to meet the mentioned observational requirements and conclude by addressing our recommendations to the Heliophysics 2024 Decadal Survey Committee.


\section{Remote-Sensing Observations} \label{sec:remote}

In terms of remote-sensing measurements, we will identify observational gaps that are crucial for a more complete understanding of CMEs and their heliospheric evolution in two main areas, namely direct imaging ({\S}\ref{subsec:directimg}) and radio probing ({\S}\ref{subsec:indirectimg}).

\subsection{Direct Imaging} \label{subsec:directimg}

As mentioned in the Introduction, the launch of the STEREO mission in 2006 represented a major advancement for CME science from a remote-sensing perspective: For the first time, the same eruption could be observed on the solar disk from more than one viewpoint (for a total of three, i.e.\ Earth plus the twin STEREOs, depending on the source region location), and three coronagraph suites were concurrently operational, providing multi-point views of the solar corona. Additionally, both STEREO spacecraft were equipped with heliospheric cameras, constantly imaging the space between the Sun and Earth (and beyond). This has enabled, for example, the development of catalogs and stereoscopic studies of CMEs imaged by both STEREO probes in white light through the solar corona \citep{vourlidas2017} and interplanetary space \citep{barnes2020}.

\begin{figure}[t!]
  \centering
    \includegraphics[width=0.99\linewidth]{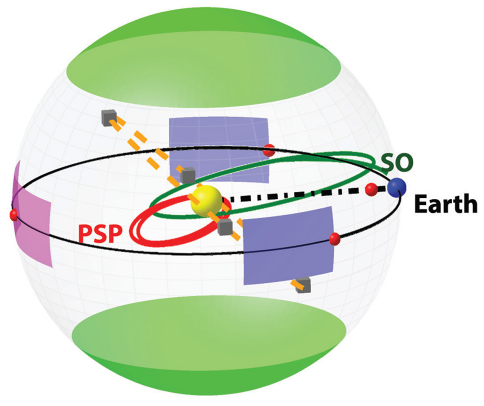}
	\caption{\footnotesize{Overview of interesting vantage points away from the Sun--Earth line. The yellow sphere indicates the Sun, while the blue sphere represents Earth. The red spheres mark the L1, L3, L4, and L5 points. The blue patches represent quadrature views (intersecting L4 and L5), the pink patch marks the far-side view, and the green patches show the polar views. `PSP' and `SO' indicate the orbits of Parker Solar Probe and Solar Orbiter, respectively. The dashed orange line shows a sample ``diamond'' orbit with four spacecraft (gray cubes) reaching heliolatitudes of $75^{\circ}$ \citep[see][]{vourlidas2018}. From \citet{gibson2018}.}}
\label{fig:viewpoints}
\end{figure}

Despite the ground-breaking progress brought by the multi-viewpoint capabilities of the STEREO mission, all solar observations to date have been characterized by one common factor, i.e.\ they have all been performed from the vicinity of the ecliptic plane. This issue will be partially addressed by Solar Orbiter, which is planned to ultimately reach a heliographic latitude of $33^{\circ}$ during its extended mission, in July 2029 \citep{muller2020}. However, even Solar Orbiter's maximum elevation with respect to the Sun's equator is significantly closer to the ecliptic plane than to the solar poles, and thus the full potential of a polar imager will not be explored. Furthermore, the ever-changing longitudinal separation between the STEREOs (and Solar Orbiter) and Earth means that the advantages and availability of these additional viewpoints are inconsistent---especially after the loss of STEREO-B in October 2014 and with STEREO-A crossing the Sun--Earth line in August 2023. The benefits of sustained (fixed) observations from quadrature and polar views for tracking the solar wind and its transient phenomena (including CMEs) have been reviewed by \citet[][see also Figure~\ref{fig:viewpoints}]{gibson2018}.

These considerations are not only valid for solar disk and coronal observations, but also for heliospheric imaging. First of all, an imager from Earth's perspective has been lacking since 2011, when the Solar Mass Ejection Imager onboard the Coriolis spacecraft was deactivated \citep[this issue will be addressed by the PUNCH mission to be launched in April 2025,][]{deforest2022}. As shown by \citet{amerstorfer2018}, a heliospheric imager can provide advantageous measurements to track CMEs that are even directed toward the observer itself (e.g., an imager at L1 would be useful to study Earth-directed CMEs). Additional heliospheric cameras from quadrature and/or polar vantage points would enable stereoscopic analysis of the evolution of CMEs long after they have left the outer corona. Moreover, the PUNCH mission will be equipped with photometric capabilities, allowing for studies of the 3D structure of the solar wind and its transients. Given the difficulties in tracking CMEs through the heliosphere even with an optimally-placed imager or a pair of imagers \citep{lugaz2010}, single-view polarization measurements can assist in overcoming the problems of stereoscopy by resolving the CME leading edge and identifying substructures within CMEs. The benefits of polarized heliospheric imaging for CME research have been discussed by \citet{deforest2016}.

Finally, a crucial capability that was not realized on the STEREO suite of remote-sensing instruments is represented by magnetographs. In addition to being able to link CMEs and their internal structure to the magnetic configuration of their source regions \citep[e.g.,][]{palmerio2017}, one of the largest sources of uncertainty in heliophysics magnetohydrodynamic (MHD) modeling is the structure and evolution of the coronal magnetic fields \citep{wiegelmann2017}. In fact, all (realistic) MHD models of the corona and inner heliosphere require the full-Sun surface-field as a boundary condition. The development of flux transport models \citep[e.g.,][]{hoeksema2020} has been largely driven by the lack of magnetograph coverage away from the Sun--Earth line. Being able to observe the full emergence, evolution, and decay of active regions on the far side of the Sun would be extremely beneficial in terms of quantification of the magnetic flux, energy, and helicity budgets associated with CME magnetic source regions \citep[e.g.,][]{vourlidas2020b}. Additionally, an increased/full coverage of solar surface magnetic field measurements would also improve model performance on CME and CME-driven shocks \citep{jin2022}.

\subsection{Radio Probing} \label{subsec:indirectimg}

An additional aspect of CME research is represented by remote-sensing observations that do not consist of ``images'' in the strictest sense. These measurements are usually realized at radio wavelengths and enable probing of different characteristics of the interplanetary medium ``from a distance'' via ground-based facilities or space-based missions.

The most widely used applications in this sense comprise observations of CME-related radio emissions in the form of noise storms and bursts \citep[see the review by][]{vourlidas2020a}. In particular, so-called Type~II radio bursts are used to track CME-driven shocks through the solar corona and inner heliosphere \citep[e.g.,][]{lara2003, gopalswamy2009b}. When radio spectra from two separate locations are available, Type~II sources can be triangulated to reconstruct shock fronts in 3D \citep[e.g.,][]{magdalenic2014}. This technique, however, can have large uncertainties that are dependent on the signal-to-noise ratio (S/N), instrument cross-calibration, and the broad angular width of the radio emission---generally restricting its use to near-Sun frequencies (MHz). Improvements in detector S/N as well as the frequency coverage and number of observing spacecraft may increase the utility of the technique for tracking ICME-driven shock propagation, especially if used in conjunction with simultaneous white-light coronal/heliospheric imagery.

\begin{figure}[t!]
  \centering
    \includegraphics[width=0.99\linewidth]{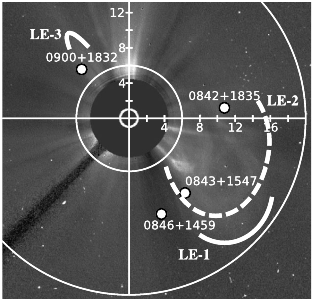}
	\caption{\footnotesize{Example of three CMEs (LE-1, LE-2, and LE-3) observed by the SOHO/LASCO/C3 coronagraph in August 2012 together with the positions of various radio galaxies that were occulted by the eruptions, allowing studies of their magnetic structure via Faraday rotation measurements. From \citet{kooi2017}.}}
\label{fig:faraday}
\end{figure}

Another way of probing CME properties at radio wavelengths is realized via Faraday rotation measurements, i.e.\ of the rotation of the polarization plane when linearly polarized radiation propagates through a magnetized plasma, e.g.\ from a CME \citep[see the review by][]{kooi2022}. In particular, Faraday rotation can be used to remotely probe CME magnetic fields \citep[e.g.,][see also Figure~\ref{fig:faraday}]{wood2020}. This technique, however, has not been explored to its fullest potential yet, since it requires the presence of a background transmitter of (at least partially) linearly polarized light, which may be either a natural radio source (e.g., a pulsar, a nebula, and/or a galaxy) or a spacecraft. This aspect poses limitations to the availability of line-of-sight observations for a given CME. Given that it is impossible to constrain the positions of galactic and extra-galactic sources, utilizing spacecraft as background radio sources may be a promising opportunity. Past observations have suffered from the lack of suitable probes in optimal positions and/or from the limited availability of ground-based telescope operations; nevertheless, some studies have successfully detected CMEs using radio signals from Pioneer~9 \citep{levy1969}, Helios \citep{bird1980}, and MESSENGER \citep{jensen2018}. These limitations and challenges could be addressed by a multi-spacecraft mission to be used as a constellation of multi-frequency background transmitters, allowing to remotely probe CME magnetic fields along multiple lines of sight.

In fact, these considerations are valid also for the interplanetary scintillation technique \citep[e.g.,][]{manoharan2010, jackson2020}, which is used to study fluctuations in the intensity of a radio source due to solar wind irregularities, allowing e.g.\ to remotely probe CME density \citep[e.g.,][]{lynch2002}.


\section{In-situ Measurements} \label{sec:insitu}

In terms of in-situ measurements, we will identify observational gaps that are crucial for a more complete understanding of CMEs and their heliospheric evolution in two main research regimes, namely those dedicated to studies of the large-scale ({\S}\ref{subsec:largescale}) and smaller-scale ({\S}\ref{subsec:smallscale}) structure of ICMEs. Note that here we consider ``small scales'' to correspond to angular separations of less than ${\sim}6^{\circ}$ at ${\sim}1$~AU, i.e.\ absolute distances of less than ${\sim}0.1$~AU.

\subsection{Large-Scale Studies} \label{subsec:largescale}

\begin{figure}[th!]
  \centering
    \includegraphics[width=0.99\linewidth]{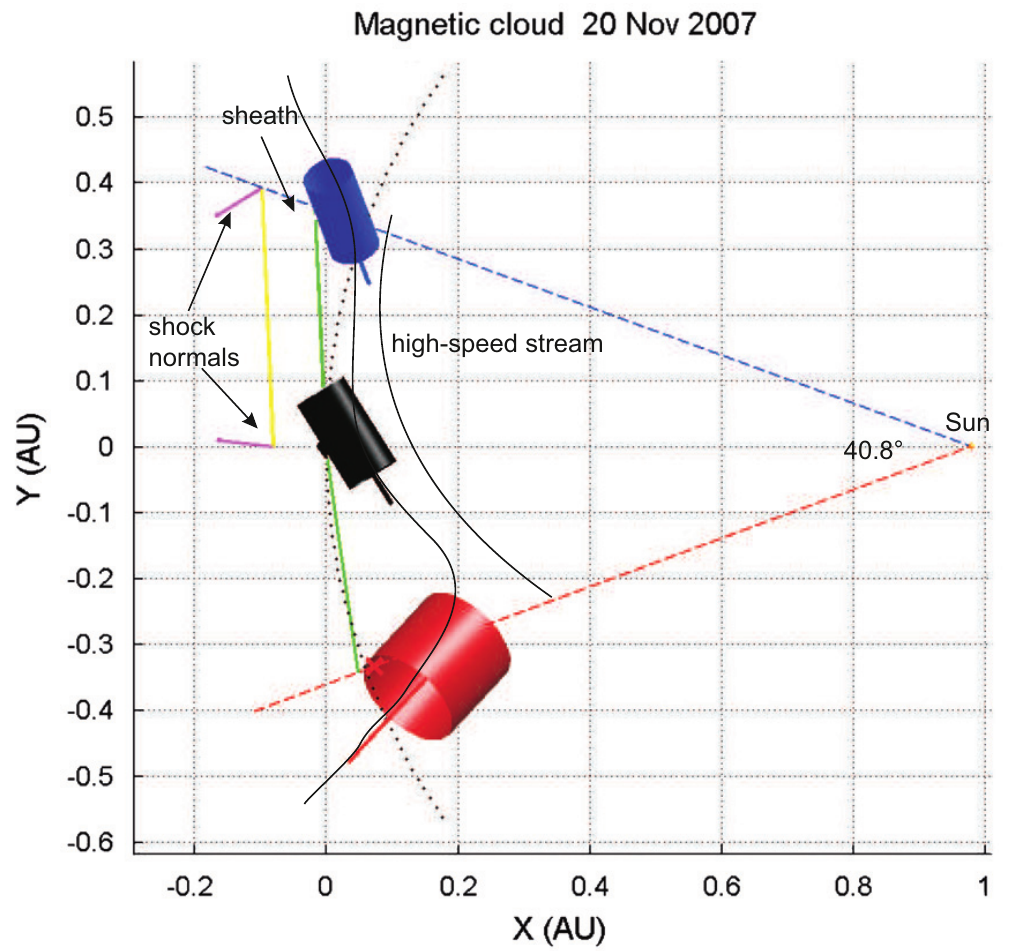}
	\caption{\footnotesize{Sketch showing the local orientation(s) of an ICME measured at 1~AU by STEREO-A, Earth, and STEREO-B in November 2007, projected onto the ecliptic plane. From \citet{farrugia2011}.}}
\label{fig:magcloud}
\end{figure}

CMEs are extremely large heliospheric structures, expanding to several times the size of the Sun already when traveling through the solar corona and reaching typical radial widths of ${\sim}3$~AU at heliocentric distances of ${\sim}15$~AU \citep{richardson2006}. Accordingly, studies that aim to address the global morphology, magnetic configuration, and plasma distribution of ICMEs tend to rely on multi-spacecraft measurements over large heliospheric distances. However, only a handful of dedicated heliophysics missions have been launched away from Earth during the past ${\sim}5$ decades (Helios, Ulysses, STEREO, Parker Solar Probe, and Solar Orbiter). Hence, the research community has been taking advantage of data from planetary missions (both during cruise phase and after orbit insertion), such as MESSENGER to Mercury, Venus Express to Venus, MAVEN to Mars, Juno to Jupiter, and Cassini to Saturn. This has led to a wealth of multi-spacecraft studies utilizing measurements ``not originally meant for CME science'' \citep[e.g.,][]{mulligan1999, nieveschinchilla2012, witasse2017, good2019, kilpua2019, davies2020, davies2021, davies2022, palmerio2021c} to accompany those based more strictly on heliophysics missions \citep[e.g.,][see also Figure~\ref{fig:magcloud}]{skoug2000, farrugia2011, winslow2021, lugaz2022}, as well as studies focused on ICMEs at other planets than Earth \citep[e.g.,][]{winslow2015,lee2017}.

Nevertheless, a handful of spacecraft (whether intended for heliophysics or planetary studies) each following its own orbit around the Sun across the vast heliosphere signifies that most multi-point CME encounters are fortuitous detections that are realized over arbitrary relative separations of the various observers, making it particularly challenging to systematically characterize CME structure and evolution. For example, if two in-situ locations are separated by ${\sim}1$~AU in heliocentric distance and ${\sim}30^{\circ}$ in heliographic longitude, it becomes nearly impossible to attribute structural and compositional differences to radial evolution, to longitudinal variations, or to both. Several works have analyzed the radial evolution of CMEs based on a few encounters characterized by near-radial alignment of the involved spacecraft \citep{good2019, vrsnak2019, salman2020}, but most multi-point events cannot rely on such rare configurations \citep[e.g.,][]{mostl2022}. Thus, measurements that are aimed to directly address multi-point observations of CMEs---e.g., via a series of probes with well-defined separations in longitude/latitude and heliocentric distance---are likely to bring outstanding advancements in the field of heliophysics.

Another important issue in CME in-situ research is represented by the absence of significant progress in many in-situ instrumentation capabilities in the past thirty years. Measurements and cadence provided by ACE and Wind for more than 25 years are often considered to be ``good enough'' for many studies, whereas higher-cadence compositional and energetic particle data, as well as advancements in radio measurements as highlighted above are necessary to make significant progress towards answering withstanding science questions.

\subsection{Smaller-Scale Studies} \label{subsec:smallscale}

\begin{figure}[th!]
  \centering
    \includegraphics[width=0.99\linewidth]{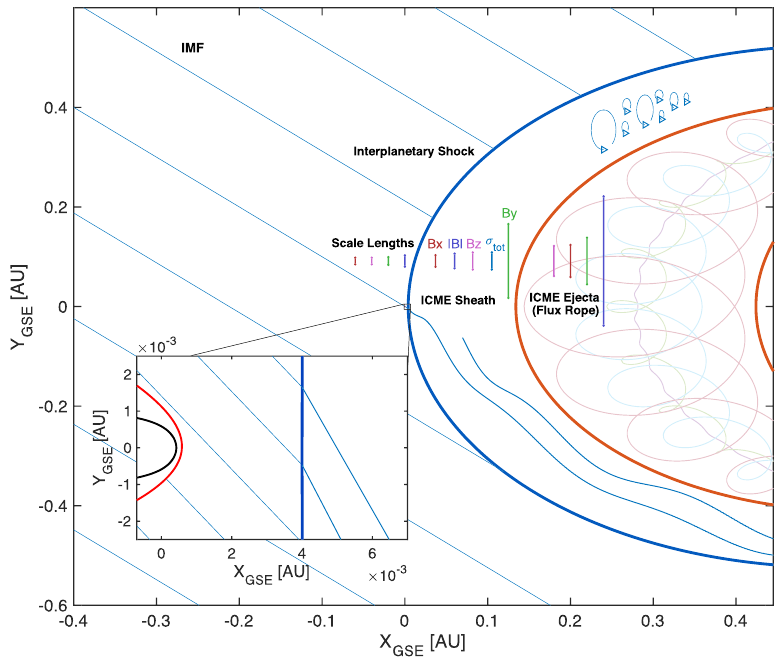}
	\caption{\footnotesize{Representative sketch of an ICME near 1~AU and the spatial scales involved. The zoomed-in inset also shows the typical size of Earth's dayside magnetosphere under nominal conditions. From \citet{alalahti2020}.}}
\label{fig:icme1au}
\end{figure}

Studies of the structure of CMEs over smaller distances (${\lesssim}0.1$~AU, corresponding to ${\lesssim}6^{\circ}$ at 1~AU) have been even less frequent. Since the STEREO mission was launched during solar minimum, only a couple of events were observed in situ before the relative separation between the twin spacecraft (and between each of them with Earth) became too large for analysis of the smaller-scale structure of CMEs \citep[e.g.,][]{liu2008, kilpua2009, mulligan2013}. Some works have taken advantage of the presence of both the ACE and Wind spacecraft near Earth. For example, \citet{mostl2008} performed an optimized flux rope reconstruction for a CME detected in November 2003 using data from both satellites, with ACE at L1 and Wind in the dawn direction closer to Earth's magnetotail. During 2000--2002, Wind performed prograde orbits, yielding separations with ACE (at L1) of ${\sim}0.01$~AU, which allowed \citet{lugaz2018} to study variations in the ejecta magnetic field for 21 ICMEs, and \citet{alalahti2020} to analyze differences in the structure of CME-driven sheaths for 29 events. These works resulted in evaluation of the typical scale lengths associated with various solar wind structures (see Figure~\ref{fig:icme1au}). For example, \citet{lugaz2018} concluded that the scale length of longitudinal magnetic coherence inside CME ejecta lies around 0.25--0.35~AU (14--$20^{\circ}$ at 1~AU) for the magnetic field magnitude, but around 0.06--0.12~AU (3--$7^{\circ}$ at 1~AU) for the magnetic field components.

These results highlight that there exists a region of the parameter space left largely unexplored, i.e.\ that corresponding to radial separations of 0.005--0.050~AU and angular separations of 1--$12^{\circ}$. This `mesoscale' region could be researched further during mid-to-late 2023, when STEREO-A will be positioned close to the Sun--Earth line, or via orbital maneuvers of the existing assets at L1, or via a dedicated multi-spacecraft mission.


\section{Improvements to Bridge Current Gaps}

The critical ``gaps'' identified in both the remote-sensing (Section~\ref{sec:remote}) and in-situ (Section~\ref{sec:insitu}) observational regimes can be addressed with a conscious, focused commitment on the part of NASA, NSF, NOAA, and other federal funding agencies to facilitate multi-point and multi-spacecraft heliophysics and planetary science missions, support for multi-probe data analysis and modeling research programs, and an innovative cross-disciplinary approach to maximize the scientific return from missions beyond their primary scientific objectives.

First of all, it is evident from the discussion in the previous sections that CME science (but more generally solar and heliospheric physics) has been largely lacking support for dedicated multi-spacecraft capabilities. The first such mission in the inner heliosphere is HelioSwarm, which was selected for flight as recently as 2022 to study turbulence in the solar wind (and is expected to launch no earlier than 2028), while spacecraft constellations have been employed for magnetospheric studies for over two decades (e.g., with Cluster, THEMIS, and MMS). Missions and assets that can enable \emph{systematic}, multi-point studies of CMEs and related phenomena in both the remote-sensing and in-situ regimes will be a prerequisite for breakthrough science and understanding to be achieved over the next decade. This much-needed progress can be realized via a multitude of observing strategies, such as coordinated spacecraft at the Lagrange L4 and L5 points (or even including the L3 point), constellations in heliocentric orbits and/or in solar polar orbits, and ``swarms'' of smaller probes (e.g., cubesats) clustered upsteam of Earth and/or other planets. A number of mission concepts relevant to these goals have been proposed over the past few years \citep[e.g.,][]{vourlidas2018, vourlidas2020b, bemporad2021, allen2022, telloni2022}.

Other practical ways to maximize science return include the incorporation of simultaneous remote-sensing \emph{and} in-situ capabilities on all heliophysics flagship missions that are set to fly through the solar wind, wherever possible. This would allow for an increased coverage of the solar atmosphere and/or extended corona in terms of imaging, and for the presence of an additional in-situ monitor scattered throughout the heliosphere. The temporal cadence of each instrument should reflect the timescales of the processes they are supposed to shed light upon: For example, coronagraph imagery from SOHO/LASCO (12-min cadence) and STEREO/COR2 (15-min cadence) is often of limited use especially in the case of fast CMEs, which can evolve dramatically over much shorter timeframes, particularly in the range ${\sim}1$--10\,$R_{\odot}$. The selection of new missions should aim to explore presently unknown regions (e.g., the solar poles) and enable novel science methodologies (e.g., using a spacecraft network as radio transmitters) but should, at the same time, guarantee continuity of the existing assets and data products.

Furthermore, synergies between heliophysics and planetary science should be facilitated and even encouraged. In addition to the multi-point CME studies enabled by planetary missions mentioned in Section~\ref{subsec:largescale}, there are e.g.\ several recent examples of novel planetary science results made with Parker Solar Probe data during its flybys of Venus \citep[e.g.,][]{collinson2021, pulupa2021, wood2022}. Another example is the community-driven, heliophysics working group formed to analyze the in-situ magnetic field and particle data being returned during BepiColombo's cruise phase \citep{mangano2021}. Such cross-disciplinary studies may be even more successful (and less fortuitous) if NASA (and other agencies) prioritize developing multi-spacecraft and multi-viewpoint capabilities at the highest levels of mission selection, funding, planning, and implementation. Two relatively straightforward action items that have the potential to maximize future scientific returns are (1) the inclusion of magnetometer, solar wind plasma, and energetic particle instruments on upcoming and future planetary science missions, and (2) to provide the funding and support for making science-quality cruise phase data readily available to the science community.


\section{Summary and Recommendations}

In this White Paper, we have outlined the current status of CME observational research and identified the current gaps that need to be filled in order to reach a more complete understanding of their internal structure, properties, and evolution. 
Our recommendations for the Heliophysics 2024--2033 Decadal Survey Committee are:
\begin{itemize}[leftmargin=2.0ex]
\item Prioritize multi-point inner heliospheric observations via coordinated spacecraft and/or constellations of probes
\item Improve simultaneous remote-sensing and in-situ coverage over the solar atmosphere/corona and the inner heliosphere
\item Observe the Sun and its environment from novel viewpoints, e.g.\ away from the ecliptic
\item Encourage CME studies that target heliospheric evolution over a range of radial and longitudinal separations
\item Address observational gaps in the CME `mesoscale' region
\item Enable higher-cadence remote and in-situ measurements relevant to CME science
\item Select new missions with the aim to explore new regions, but without losing continuity of the existing assets
\item Include heliophysics-relevant instrumentation on planetary missions, to maximize cross-disciplinary studies and science return
\end{itemize}

\clearpage


\bibliographystyle{wp_bibstyle}
\bibliography{bibliography}

\end{document}